\documentclass[amsmath,amssymb,twocolumn,pre,showpacs]{revtex4}

\usepackage{graphicx}

\newcommand*{\dd}{\mathrm{d}}
\newcommand*{\ind}[1]{_{\text{#1}}}
\newcommand*{\hoch}[1]{^{\text{#1}}}
\newcommand*{\ket}[1]{\mathinner{\lvert#1\rangle}}
\newcommand*{\bra}[1]{\mathinner{\langle#1\rvert}}
\newcommand*{\ketbra}[2]{\ket{#1}\!\bra{#2}}
\newcommand*{\expec}[1]{\mathinner{\langle#1\rangle}}
\newcommand*{\abs}[1]{\lvert#1\rvert}
\newcommand*{\op}{\hat}
\newcommand*{\sop}{\op\sigma}
\newcommand*{\pop}[2]{\frac{\partial #1}{\partial #2}}
\newcommand*{\Tr}{\operatorname{Tr}}

\newcommand*{\cs}{S}
\newcommand*{\env}{C}

\begin{document}

\title{Control of Local Relaxation Behavior in Closed Bipartite
  Quantum Systems}

\author{Harry Schmidt}
\affiliation{Institut f\"ur
  Theoretische Physik 1, Universit\"at Stuttgart}

\author{G\"unter Mahler} 
\affiliation{Institut f\"ur
  Theoretische Physik 1, Universit\"at Stuttgart}

\begin{abstract}
  We investigate the decoherence of a spin~$1/2$ subsystem weakly
  coupled to an environment of many spins~$1/2$ with and without
  mutual coupling.  The total system is closed, its state is pure and
  evolves under Schr\"odinger dynamics.  Nevertheless, the considered
  spin typically reaches a quasi-stationary equilibrium state.

  Here we show that this state depends strongly on the coupling to the
  environment on the one hand and on the coupling within the
  environmental spins on the other.  In particular we focus on spin
  star and spin ring-star geometries to investigate the effect of
  intra-environmental coupling on the central spin.  By changing the
  spectrum of the environment its effect as a bath on the central spin
  is changed also and may even be adjustable to some degree.  We find
  that the relaxation behavior is related to the distribution of the
  energy eigenstates of the total system.

  For each of these relaxation modes there is a dual mode for which
  the resulting subsystem approaches an inverted state occupation
  probability (negative temperature).
\end{abstract}


\maketitle

\section{Introduction}
\label{sec:intro}

In a composite but closed quantum system in which a smaller ``central
system'' is weakly coupled to a larger ``environment'', most of the
(pure) states of the total system for a given energy (and possibly
some additional constraints) exhibit properties of thermal equilibrium
states with respect to the smaller part
\cite{mahler:quant_therm,gemmer.2003}, i.e.~there exists a so-called
dominant region in Hilbert space in which the entropy of the central
system is close to its maximum value under the given constraints.  For
a completely ``unstructured'' coupling to the environment, the state
of the central system, starting away from equilibrium, shows
decoherence and thermalization and, ultimately, a quasi-stationary
equilibrium situation is reached which is determined only by the
spectrum of the environment \cite{borowski.2003}.

In this paper we investigate to what extent structured systems will
also exhibit this thermalization behavior.  We focus, in particular,
on a single central spin~$1/2$ particle coupled to a relatively small
environment of spin~$1/2$ particles.  Recently, the properties of spin
systems of different structure (rings, stars and others) have been
subject of extensive interest.  A lot of work has been done on the
question of
entanglement~\cite{briegel.2001,oconnor.2001,wang.2002,hutton.2004,hutton.2004a,vidal.2003},
their relaxation behavior has been
addressed~\cite{breuer.2004,lages.2004} and various techniques were
suggested to make any spin interact with any other
spin~\cite{imamoglu.1999,makhlin.1999}.

Here we show that by choosing different types of coupling between the
central system and the environment on the one hand and within the
environment on the other the equilibrium state finally reached can be
controlled.  There is, in particular, a qualitative difference between
the relaxation dynamics and the equilibrium of the central spin for
interacting and noninteracting environments.  We relate this
equilibrium to the spectral structure of the total system.

\newpage

\section{Spectral Temperature}
\label{sec:spectral-temperature}

Consider a weakly coupled bipartite quantum system consisting of a
small system~\cs\ and a large ``container'' or environment~\env\ with
the Hamiltonian
\begin{equation}
  \label{eq:1}
  \op H = \op H\hoch{\cs} + \op H\hoch{\env} +
  \alpha \op H\hoch{int}, \quad \expec{\alpha \op H\hoch{int}} \ll
  \expec{\op H\hoch{\cs}}, \expec{\op H\hoch{\env}}.
\end{equation}
The behavior of the system is completely described by a Hilbert space
vector~$\ket{\psi(t)}$ and its evolution under the Schr\"odinger
equation.  Under weak constraints, the state of~\cs\ after relaxation
is determined by Jaynes' principle, taking energy as the only relevant
observable~\cite{mahler:quant_therm,gemmer.2003}.  The temperature of
the respective canonical state can be predicted from the degeneracy
structure of the environment alone.

\begin{figure}[bp]
  \centering
  \includegraphics{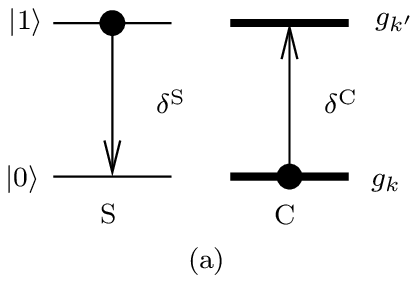}%
  \hspace{2em}
  \includegraphics{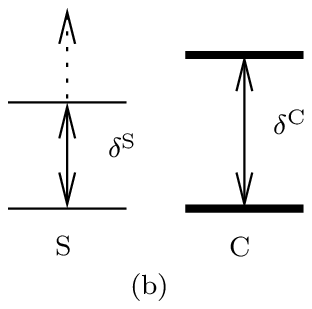}
  \caption{Two-level system~\cs\ in contact with environment~\env\
    consisting of two highly degenerate levels~$k$,~$k'$ with
    degeneracies~$g_k$ and~$g_{k'}$, respectively.  (a)~Resonance,
    $\delta\hoch{\cs}=\delta\hoch{\env}$, showing the
    state~$\ket{1}\hoch{\cs}\otimes\ket{k,m}\hoch{\env}$ for which~\cs\
    evolves into a thermal state. (b)~Slightly detuned energy
    splitting between~\cs\ and~\env: Population relaxation in~\cs\ is
    prohibited by energy conservation,~\env~no longer acts as an
    effective heat bath.}
  \label{fig:2-lvl-s_2-lvl-c}
\end{figure}

Fig.~\ref{fig:2-lvl-s_2-lvl-c}(a) shows a two-level system in
resonant contact with an environment consisting of two ``energy
bands''~$k$,~$k'$ of degeneracies~$g_k$ and~$g_{k'}$, respectively,
($g_k,g_{k'}\gg1$) in a non equilibrium (initial) state.  In
equilibrium, the state of the total system is expected to be
distributed homogeneously over the whole Hilbert space on average, the
time-averaged reduced state operator of~\cs\ is thus given
by~\cite{mahler:quant_therm}
\begin{equation}
  \label{eq:2}
  \op\rho\hoch{\cs} = \frac{1}{g_k+g_{k'}} \bigl( g_{k'} \ketbra{0}{0} + g_k
  \ketbra{1}{1} \bigr)
\end{equation}
which can be interpreted as a canonical state operator with inverse
temperature
\begin{equation}
  \label{eq:3}
  \beta\hoch{\cs} = \frac{1}{k\ind{B}T} = 
  \frac{1}{\Delta E} \ln \frac{g_{k'}}{g_k}.
\end{equation}
In general, the inverse equilibrium temperature imparted on~\cs\ for
participating energy levels~$E$ of the environment~\env\ can be
defined as the derivative of the logarithm of the
degeneracy~$g\hoch{\env}(E)$ with respect to energy,
\begin{equation}
  \label{eq:4}
  \beta\hoch{S} = \pop{\ln(g\hoch{\env}(E))}{E}.
\end{equation}
Note that this is a spectral property only and does not imply that the
environment was in a thermal state; in fact, under the condition of
Schr\"odinger dynamics for the total system it is far away from such a
state.
If the degeneracy of environmental states increases
exponentially with energy, $\beta\hoch{\cs}$ is independent of the
initial state of the total system~($\text{\cs}+\text{\env}$).  In this
case also systems~\cs\ of higher dimensional Hilbert space will evolve
into canonical states.

Equations~(\ref{eq:2}--\ref{eq:4}) are valid only if energy exchange
between~\cs\ and~\env\ is possible, i.e.~if transitions in both
subsystems are in resonance.  Under more general conditions it is far
from clear what numbers should count as the pertinent
degeneracies~$g_k$.  One may expect that the action of the environment
as an effective heat bath could even be turned on and off by tuning
the level splitting in either system as shown in
Fig.~\ref{fig:2-lvl-s_2-lvl-c}(b).  In case of entirely detuned
transitions the environment would still induce decoherence, i.e.~phase
relaxation in~\cs, thus effectively acting as a microcanonical
environment.

At first glance one would expect both parts of the system to be in
resonance if the detuning~$\abs{\delta\hoch{\cs}-\delta\hoch{\env}}$
was smaller than the width of the involved bands in~\env, i.e.~it best
should be zero.  However, if the free Hamiltonian of~\cs\ is
renormalized by the environment it is possible that the transitions
within the central system and the environment become somewhat off
resonant for vanishing detuning.  In this paper we focus on values of
the detuning with maximal overlap, i.e.~we
choose~$\delta\hoch{\cs}-\delta\hoch{\env}$ such that the equilibrium
state was as close as possible to the expected one from
equation~\eqref{eq:3}.  However, for part of the following discussion
the precise value of the detuning is not very important,
and~$\delta\hoch{\cs}$ could have been chosen equal
to~$\delta\hoch{\env}$.

\section{Choice of Environmental Spectrum}
\label{sec:choice-envir-spectr}

To use the quantum environment as an effective heat bath for the
central system~\cs\ with adjustable temperature, though, the
degeneracy of the environment must not scale exponentially with
energy.  Then the equilibrium temperature of the subsystem depends on
the initial state of the environment via equation~\eqref{eq:4} and can
be adjusted by adequately preparing the initial environmental state.
For the equilibrium state of the system~\cs\ to be sufficiently
independent of its initial state, the spectral temperature should vary
only little over adjacent energy levels, though.

\begin{figure}
  \centering
  \includegraphics{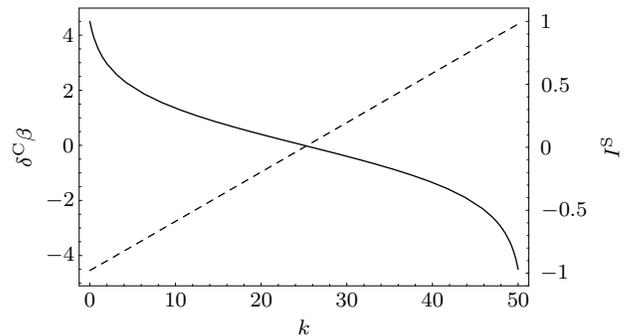}
  \caption{Solid line: Inverse equilibrium
    temperature~$\beta\hoch{\cs}$ of~\cs\ in units
    of~$\delta\hoch{\env}$ depending on the involved energy band~$k$
    of \env.  Dashed line: Corresponding population
    inversion~$I\hoch{\cs}$ in equilibrium.  The results shown are for
    $N=50$.}
  \label{fig:beta-vs-k}
\end{figure}

A possible implementation of these requirements is a degeneracy
varying binomially with energy \cite{diu:statistische_physik},
\begin{equation*}
  g_k = \binom{N}{k}, \quad E_k = k\delta\hoch{\env},
\end{equation*}
where we assume resonance of~\cs\ with adjacent energy bands~$k$
and~$k+1$, $0\le k,k+1\le N$.  The band splitting~$\delta\hoch{\env}$
will be taken as a convenient energy scale.  For~$N=50$
Fig.~\ref{fig:beta-vs-k} shows the equilibrium temperature of~\cs\ as
a function of the ``band index''~$k$ of the environmental initial
state and the corresponding population inversion~$I\hoch{\cs} =
\bra{1}\op\rho\hoch{\cs}\ket{1} - \bra{0}\op\rho\hoch{\cs}\ket{0}$.
The inversion of the state~\eqref{eq:2} depends on~$k$ and is given by
\begin{equation}
  \label{eq:5}
  I\hoch{\cs} = \expec{\op\sigma_z}  = \frac{g_k - g_{k+1}}{g_k + g_{k+1}}.
\end{equation}
For sufficiently large~$N$ and~$0\le k\le N$, the
temperature~$T\propto 1/\beta\hoch{\cs}$ varies over a wide range of
positive, infinite and even negative values, although locally,
i.e.~around a given energy band, the deviations from an exponential
degeneracy-structure are still small, except for very small or
large~$k$.

\section{Spin Environments}
\label{sec:spin-environments}

An environment consisting of $N$ spin~$1/2$ particles with dominant
Zeeman-terms
\begin{equation*}
  \op H\hoch{\env} = \frac{\delta\hoch{\env}}{2} \op J_z + \op
  H\hoch{\env\env},\quad
  \op J_z = \sum_{\nu=1}^N \op\sigma_z^{(\nu)}
\end{equation*}
(mutual interaction~$\op H\hoch{\env\env}$ within~\env\ small) has the
desired binomial degeneracy, $\op\sigma_z^{(\nu)}$ denotes the
Pauli-operator of the $\nu$\textsuperscript{th} particle.  To this
environment we couple the central spin with the local Hamiltonian $\op
H\hoch{\cs} = \frac{\delta\hoch{\cs}}{2}\op\sigma_z$.

We will now consider several different types of
interaction~$\op{H}\hoch{int}$ between~\cs\ and~\env.  To test whether
the system shows thermalization for a given~$\op H\hoch{int}$, we
study the Schr\"odinger time evolution of the initial product state
depicted in Fig.~\ref{fig:2-lvl-s_2-lvl-c}(a),
\begin{equation}
  \label{eq:6}
  \ket{\psi\hoch{\cs$+$\env}(t=0)} =
  \ket{1}\hoch{\cs} \otimes \ket{k,m}\hoch{\env},
\end{equation}
where~$k$ denotes an energy band in the environment and~$m$ indicates
one state within this band.  If the coupling is weak
and~$\delta\hoch{\cs}\approx\delta\hoch{\env}$ energy conservation
restricts the dynamics mainly to a subspace of Hilbert space dimension
\begin{equation}
  \label{eq:7}
  d\ind{H} = g_k + g_{k+1} = \binom{N}{k}+\binom{N}{k+1}
\end{equation}
spanned by the states~$\bigl\{\ket{1}\hoch{\cs}\otimes\ket{k,m}\hoch{\env},
\ket{0}\hoch{\cs}\otimes\ket{k+1,m'}\hoch{\env}\bigr\}$
($m=1,2,\dots,g_k$ and $m'=1,2,\dots,g_{k+1}$).  After a suitable
thermalization time the reduced state operator of the central
system~$\op{\rho}\hoch{\cs}$ should be close to the one given
by~\eqref{eq:2}.

The subspace of Hilbert space spanned by the~$g_k+g_{k+1}$ state
vectors~$\{\ket{1}\hoch{\cs}\otimes\ket{k,m}\hoch{\env},\ket{0}\hoch{\cs}\otimes\ket{k+1,m'}\hoch{\env}\}$
will be called the ``accessible subspace'' throughout this paper.
Nevertheless, other states have to be included, if one is interested
in the off-diagonal elements of the reduced state of~\cs.  The
accessible subspace is to be distinguished from the subspace that is
actually ``involved'' in the dynamics from a given initial state via
Schr\"odinger time evolution.  This ``involved'' subspace usually is a
subset of the ``accessible'' one.

\subsection{Random Interaction}
\label{sec:random-interaction}

To test the expectations of the previous sections we first
take~$\op{H}\hoch{\env\env}=0$ and~$\op{H}\hoch{int}$ as a hermitian
random matrix form the Gaussian unitary ensemble
in~$2^{N+1}$-dimensional Hilbert space with the
distribution~$w(\op{H}\hoch{int})=C\exp\bigl(-A\Tr\{\op{H}\hoch{int}\}^2\bigr)$~\cite{haake:quant-chaos}.
It has been shown previously~\cite{borowski.2003} that an initial
product state evolves into the desired canonical state under the time
evolution generated by the given Hamiltonian.  This is not surprising,
since, in general, the complete information about the environment
being constructed from spin~$1/2$ particles is lost and energy remains
the only conserved quantity.  Most states in the accessible region of
Hilbert space are equilibrium states with respect to the central
system.

\begin{figure}[tp]
  \centering
  \includegraphics{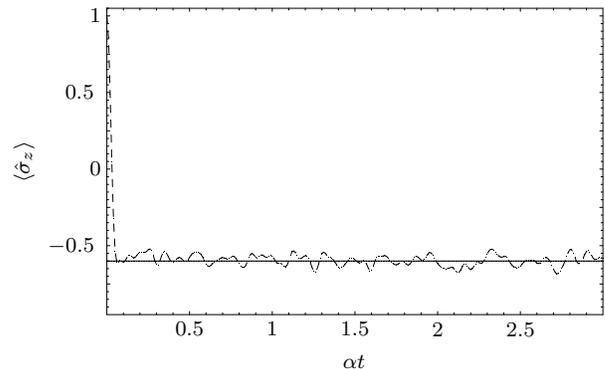}
  \caption{$z$-component of the Bloch vector of~\cs\ for initial
    state~\eqref{eq:6} ($N=14$, $k=2$, $g_3=\binom{14}{3}$,
    $g_2=\binom{14}{2}$) and random perturbations as a function of
    time~$t$.  The horizontal line indicates the expected mean value
    $I\hoch{\cs}=\expec{\sop_z}=-3/5$ (see~\eqref{eq:5}).}
  \label{fig:sigma-z_t_random}
\end{figure}

Fig.~\ref{fig:sigma-z_t_random} shows the time evolution of the
$z$-com\-po\-nent~$\expec{\sop_z}=\Tr\{\sop_z\,\op\rho\hoch{\cs}(t)\}$
of the Bloch vector for $N=14$~particles in~\env\ and initial
state~\eqref{eq:6} with~$k=2$, $\delta\hoch{\cs}=\delta\hoch{\env}$
and very weak coupling~$\alpha=\delta\hoch{\env}/5000$.  The
accessible subspace of Hilbert space is of
dimension~$d\ind{H}=g_2+g_3=455$.  Regardless of the small number of
dimensions one can observe relaxation to the expected mean value
indicated by the horizontal line.  The average over all times
\begin{equation*}
  \overline{\expec{\sop_z}} = \lim_{T\rightarrow\infty} 
  \frac{1}{T}
  \int_{-T/2}^{T/2} \Tr\{\sop_z\,\op\rho\hoch{\cs}(t)\}\,\dd t
\end{equation*}
is~$\overline{\expec{\sop_z}}=-0.598$ for the simulation of
Fig.~\ref{fig:sigma-z_t_random}.  Considering the small number of
spins this is remarkably, but not unexpectedly, close to the expected
average~$-3/5 = -0.6$.

A change of the coupling strength~$\alpha$ alone affects only the timescale
of the dynamics, provided the coupling stays weak, since
all energy eigenstates of the uncoupled system in the accessible
subspace are degenerate.  The detuning can be changed slightly without
disturbing the system qualitatively as long as the introduced level
splitting is considerably smaller than the interaction energy.

\subsection{Spin-Star Configuration}
\label{sec:spin-star-conf}

The situation is quite different if the interaction is structured.
Neglecting intra-environmental interaction
(i.e.~$\op{H}\hoch{\env\env}=0$), the most general Hamiltonian for
coupling the central spin to each spin in the environment is
\begin{equation}
  \label{eq:8}
  \op H\hoch{int} = \sum_{i,j=1}^3 \sum_{\nu=1}^N \gamma_{ij}^{(\nu)}\,
  \op\sigma_i \otimes \op\sigma_j^{(\nu)}.
\end{equation}
Even highly symmetric realizations of this class like the Heisenberg
$XX$ interaction~\cite{lieb.1961} show dissipation and decoherence
with respect to~\cs\ for highly mixed initial
states~\cite{breuer.2004}.

\begin{figure}[tp]
  \centering
  \includegraphics{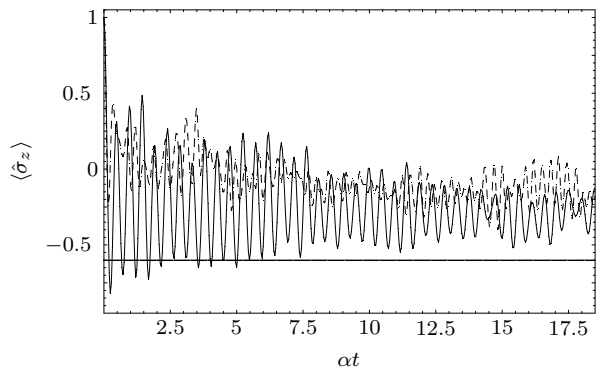}
  \caption{$z$-component of the Bloch vector of \cs\ for initial
    state~\eqref{eq:6} ($N=14$, $k=2$) for two different choices of
    $\op{H}\hoch{int}$:  Random two-body interactions,
    $\delta\hoch{\cs}=1.00082\delta\hoch{\env}$ (solid line),
    $\delta\hoch{\cs}=0.9992\delta\hoch{\env}$ (dashed line).}
  \label{fig:sigma-z_t_rand-2-body}
\end{figure}

However, this is no longer true for the pure initial
state~\eqref{eq:6}.  Fig.~\ref{fig:sigma-z_t_rand-2-body} shows the
time evolution of~$\expec{\sop_z}$ of~\cs\ for~$N=14$ particles in the
environment and~$\delta\hoch{\cs}\approx\delta\hoch{\env}=5000\alpha$.
The coefficients~$\gamma_{ij}^{(\nu)}$ have been chosen randomly from
a normalized Gaussian distribution.  Here oscillations with a
significantly larger amplitude than in the previous section are
observed because a smaller fraction of Hilbert space is involved in
the time evolution.  Also the time-averaged~$\expec{\sop_z}$ differs
considerably from the expected value indicated by the horizontal line.
This average over all times is~$\overline{\expec{\sop_z}} = -0.0976$
for the dashed and~$\overline{\expec{\sop_z}} = -0.261$ for the solid
line.

As stated earlier,~$\delta\hoch{\cs}-\delta\hoch{\env}$ was chosen to
``optimize'' the relaxation behavior with respect to approximating
state~\eqref{eq:2}.  Any other value than that used would lead to an
average reduced state of~\cs\ even further away from the expected one.

\subsection{Ring-Star Configuration}
\label{sec:spin-wheel-conf}

\begin{figure}[tp]
  \centering
  \includegraphics{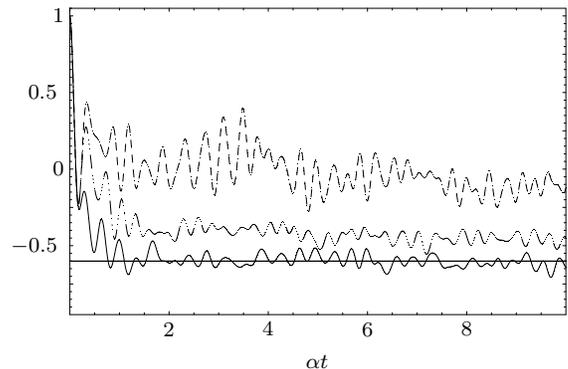}
  \caption{$z$-component of the Bloch vector for initial
    state~\eqref{eq:6} ($N=14$, $k=2$) with stronger (solid line),
    weaker (dashed-dotted line) and without (dashed line)
    intra-environmental coupling.  See the text for the respective
    values of~$\gamma$.}
  \label{fig:sigma-z_t_rand2bd_xx}
\end{figure}

Now we extend the spin-star model by introducing mutual interaction of
the environmental spins.  Next neighbor coupling in the form of the
quantum Ising chain,
\begin{equation*}
  \op H\hoch{\env\env} = \gamma \sum_{\nu=1}^{N} \sop_x^{(\nu)} 
  \otimes \sop_x^{(\nu+1)},
\end{equation*}
is added to the system discussed in the last section (assuming
periodic boundary conditions, i.e.~$\sop_x^{(N+1)} = \sop_x^{(1)}$).
Fig.~\ref{fig:sigma-z_t_rand2bd_xx} compares the time evolution of the
spin-star model as discussed in the last section (dashed line, same as
dashed line in Fig.~\ref{fig:sigma-z_t_rand-2-body}) with the same
system to which this~$\op H\hoch{\env\env}$ has been added (solid line
with~$\gamma=3\alpha$ and dashed-dotted line with~$\gamma=3\alpha/2$).
The oscillation amplitude is reduced considerably and the time
averaged value is closer to the expected one.  Here we
get~$\overline{\expec{\sop_z}}=-0.591$ for stronger
and~$\overline{\expec{\sop_z}}=-0.432$ for weaker additional
intra-environmental interaction.

\begin{figure}[tp]
  \centering
  \includegraphics{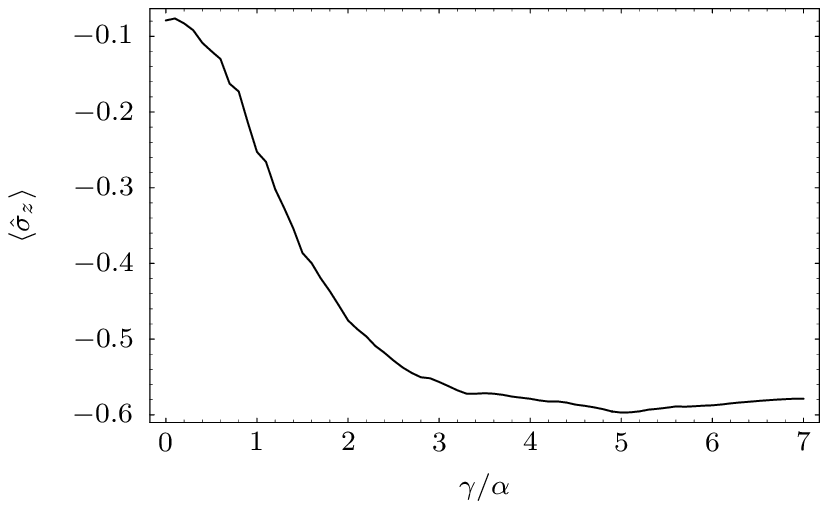}
  \caption{$\overline{\expec{\sop_z}}$ for different values
    of~$\gamma$.  Initial state~\eqref{eq:6} ($N=14$, $k=2$),
    $\delta\hoch{\cs}=\delta\hoch{\env}$.}
  \label{fig:mean-sigma-z_gamma}
\end{figure}

The relaxation dynamics and equilibrium state of~\cs\ strongly depend
on the coupling within the environment.  By increasing~$\gamma$ from
zero the amplitude of the oscillations is reduced and the average
value of~$\expec{\sop_z}$ approaches its expected value.
This is explicitly shown in Fig.~\ref{fig:mean-sigma-z_gamma}, which
also shows that the precise value
of~$\delta\hoch{\cs}-\delta\hoch{\env}$ is not too important here.  We
have found similar behavior for $XY\!$- and Heisenberg-type
interaction.  Ising-type interaction~($\sop_z\otimes\sop_z$) is
somewhat different: By carefully adjusting the parameters the system
approaches an average state close to the expected one, yet the
fluctuations are not decreased considerably.  It is important to note
that interactions involving~$\sop_z$ change the effective band
splitting in the environment so that the value of the detuning becomes
significant and has to be adjusted accordingly.

One may imagine to switch on or off
the interaction~$\op H\hoch{\env\env}$ or to tune its strength
dynamically (via ``refocusing'' pulses, see
e.g.~\cite{stollsteimer.2001}).  In this way the environment would
become adjustable with respect to its effect on the given central
spin.  This is a phenomenon which in ``classical'' thermostatistics
could hardly have been anticipated.  Furthermore, this shows how to
control~\cs\  entirely by a specific modification of its environment.

\begin{figure}
  \centering
  \includegraphics{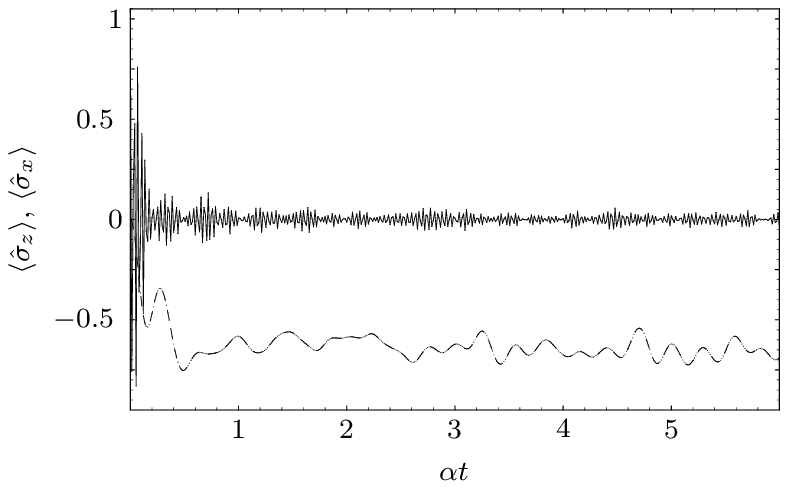}
  \caption{$x$- and $z$-components of the Bloch vector for an initial
    superposition~$(\ket{0}+\ket{1})/\sqrt{2}$ of~\cs\ as functions of
    time.  Solid line:~$\expec{\sop_x}$, dashed
    line:~$\expec{\sop_z}$.}
  \label{fig:sigma-x-z-decoh}
\end{figure}

Finally we note that in addition to showing dissipation (relaxation of
the diagonal of~$\rho\hoch{\cs}$) superpositions in~\cs\ are decohered
(the off-diagonal elements are zero on average), see
Fig.~\ref{fig:sigma-x-z-decoh}.

\section{Spectral Properties}
\label{sec:spectral-properties}

Finally, we try to attribute the various types of relaxation behavior
of these bipartite quantum systems to the distribution of the energy
eigenstates in Hilbert space or, more precisely, to the distribution
of certain properties of the energy eigenstates.  Let
$\ket{\varepsilon_n}$ be an energy eigenstate of the total system
defined by~\eqref{eq:1}.  Then
\begin{equation*}
  \op\varepsilon_n\hoch{\cs} =
  \Tr\ind{\env}\{\ketbra{\varepsilon_n}{\varepsilon_n}\}
\end{equation*}
is the corresponding reduced state of~\cs.  Since energy conservation
restricts the dynamics to some Hilbert subspace as discussed in
section~\ref{sec:spin-environments} (the accessible subspace), we only
take the energy eigenstates of this subspace into account in the
following discussion.

\begin{figure}
  \centering
  \includegraphics{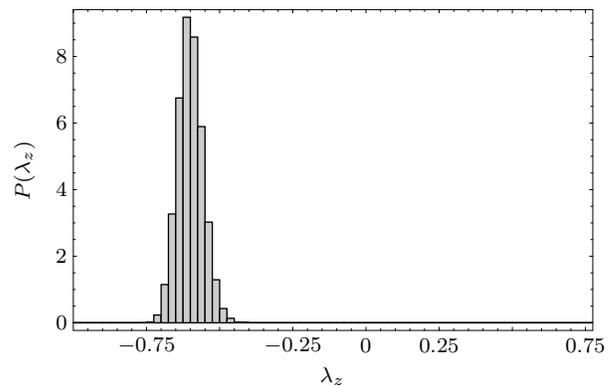}
  \caption{Distribution of the $z$-component of the reduced state
    operator corresponding to the energy-eigenstates for a completely
    random perturbation as discussed in
    section~\ref{sec:random-interaction}, 14~particles in \env.}
  \label{fig:EvecRed_random_100}
\end{figure}

\begin{figure}
  \centering
  \includegraphics{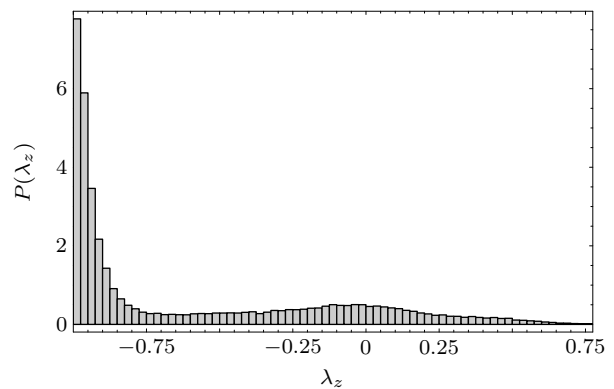}
  \caption{Distribution of the $z$-component of the reduced state
    operator corresponding to the energy-eigenstates for the spin-star
    configuration discussed in section~\ref{sec:spin-star-conf},
    14~particles in \env.}
  \label{fig:EvecRed_rand-2-body_100}
\end{figure}

\begin{figure}
  \centering
  \includegraphics{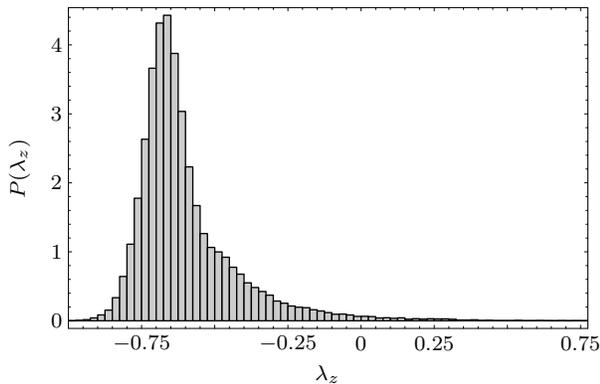}
  \caption{Distribution of the $z$-component of the reduced state
    operator corresponding to the energy-eigenstates for the ring-star
    configurations discussed in section~\ref{sec:spin-wheel-conf},
    14~particles in \env.}
  \label{fig:EvecRed_rand2bdy_XX_100}
\end{figure}

An interesting property of the~$\op\varepsilon_n\hoch{\cs}$ is
the~$z$-component of their Bloch
vectors~$\lambda_{z,n}=\Tr\{\sop_z\op\varepsilon_n\hoch{\cs}\}$.  The
average of all~$\lambda_{z,n}$ of a given Hamiltonian in the discussed
energy band is approximately given by the expected value for the
equilibrium~\eqref{eq:5}.  Figures~\ref{fig:EvecRed_random_100}
to~\ref{fig:EvecRed_rand2bdy_XX_100} show the distributions of
the~$\lambda_{z,n}$ for each of the models ($N=14$ environmental
spins) discussed in the previous section over the ensemble of the
respective random Hamiltonians.  These distributions have been
calculated numerically for 100~different randomly chosen realizations
of each model, i.e.~45500~values in total for 14~environmental spins
(as in section~\ref{sec:spin-environments}).

Fig.~\ref{fig:EvecRed_random_100} shows the distribution of the values
of~$\lambda_z$ for the complete random interaction discussed in
section~\ref{sec:random-interaction}.  Because the eigenvectors of
these Hamiltonians are homogeneously distributed over the unit sphere
in the accessible subspace, the~$\lambda_z$ are narrowly peaked
around~$-0.6$~\cite{schmidt.2005-UP2}.  Due to the homogeneous
distribution of the eigenvectors one expects most of them to
contribute to the dynamics with approximately equal weight and
therefore~$\overline{\expec{\sop_z}}$ is expected to be close to the
average~$\lambda_z$.

For the spin-star configuration shown in
Fig.~\ref{fig:EvecRed_rand-2-body_100} the situation is quite
different.  Although the average over all~$\lambda_z$ is still~$-0.6$,
there is a strong peak at~$\lambda_z=-1$.  All the energy
eigenstates corresponding to the region~$\lambda_z \approx -1$ are
approximately product states of the form
\begin{equation*}
  \ket{\varepsilon_n} \approx \ket{0}\hoch{\cs}\otimes
  \ket{\chi}\hoch{\env}
\end{equation*}
with some state $\ket{\chi}\hoch{\env}$ of the environment (a
corresponding peak for eigenstates of the
form~$\ket{\varepsilon_n'}\approx\ket{1}\hoch{\cs}\otimes\ket{\chi}\hoch{\env}$
is missing here since this state is not part of the accessible
subspace).  For the initial condition given by~\eqref{eq:6} these
states do not contribute to the dynamics, therefore the subspace
``involved'' in the dynamics is only a small fraction of the
``accessible'' subspace, leading to larger fluctuations in the time
evolution.  Also the average~$\lambda_z$ of the contributing states is
larger than~$-0.6$, therefore the equilibrium temperature differs from
its value for a completely random perturbation.  The fluctuations
could be decreased simply by using a larger environment.  However, the
average~$\lambda_z$ of the contributing energy eigenstates, and
therefore also the average equilibrium state would not be affected.

The shape of the distribution can be changed by introducing
intra-environmental coupling.  For the types of interaction discussed
in the previous section the peak is shifted to larger values
of~$\lambda_z$ for increasing coupling strength,
Fig.~\ref{fig:EvecRed_rand2bdy_XX_100} shows the result for
$\gamma=3\alpha$.  The distribution is considerably broader than in
Fig.~\ref{fig:EvecRed_random_100}, while the peak is again close
to~$\lambda_z=-0.6$ and the ``involved'' subspace is enlarged.  Since
no large fraction of the energy eigenstates is excluded here by the
choice of the initial state, most of the~$\lambda_z$ contribute and
one expects~$\overline{\expec{\sop_z}}$ to be again close to the
average~$\lambda_z$.

\section{``Thermal Duality'': Relaxation to Negative Temperature States}
\label{sec:negat-absol-temp}

For~$k>N/2$ the degeneracy of the environment decreases with energy,
$\partial\ln(g\hoch{\env}(E))/\partial{E}<0$, and therefore~\env\ can
act as an environment imparting a negative absolute temperature to
the central spin.

We have shown that the environment leads for the considered subsystem
to Schr\"odinger relaxation approaching a quasi-stationary state with
positive temperature (for~$k<N/2$).  These scenarios may be considered
as generic, if simplified, models for what we typically observe in
real life.  The environments with a finite number of states allow also
for a different class of relaxation modes: As a typical consequence of
a quantum approach to thermodynamics we find relaxation towards
negative temperature states as we replace the accessible
environment band index~$k$ by~$N-k$ (the relaxation behavior of the
system is fully symmetrical with respect to the transformation
$\ket{s}\otimes\ket{k,m}\rightarrow\ket{\bar{s}}\otimes\ket{N-k,m}$,
where $\bar s$ means logical negation)!  Contrary to inversion due to
optical pumping, these dual states would be dynamically stable just as
their positive temperature counterparts.  Note, again, that the
environment is not a heat bath of negative temperature; in fact, it is
far from any canonical state.

Since the population of the energy levels of~\cs\
at~$T<0$ are inverted ($I\hoch{\cs}>0$), negative temperatures are
actually ``hotter'' than positive ones.  Spin systems and
thermodynamics at~$T<0$ have been considered before, see
e.g.~\cite{ramsey.1956,landsberg.1977}.  In this regime several
statements of thermodynamics, like the second law, have to be
reformulated.

Here we show relaxation of~\cs\ to a state of negative absolute
temperature.  The situation is the same as discussed in
sections~\ref{sec:choice-envir-spectr}
and~\ref{sec:spin-environments}, but now we
use~$\ket{0\hoch{\cs};k,m\hoch{\env}}$
and~$\ket{1\hoch{\cs};k',m\hoch{\env}}$ with~$k=12=N-2$,~$k'=11=N-3$
(instead of~$k=2$,~$k'=3$) as the accessible subspace for the
Schr\"odinger time evolution (see Fig.~\ref{fig:beta-vs-k}).

\begin{figure}
  \centering
  \includegraphics{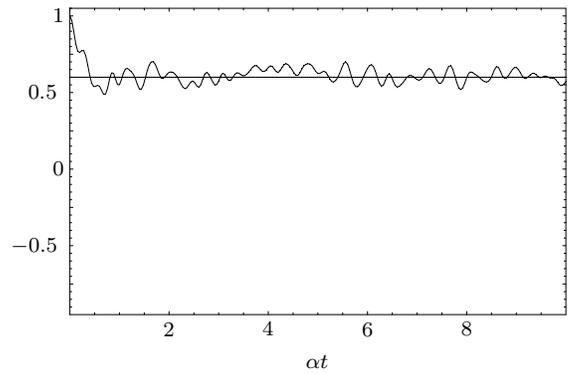}
  \caption{$z$-component of the Bloch vector for the ring-star
    geometry of section~\ref{sec:spin-wheel-conf} with initial
    state~\eqref{eq:6} ($N=14$, $k=11$).}
  \label{fig:sigma-z_t_rand2bd_xx_N14-negT}
\end{figure}

Fig.~\ref{fig:sigma-z_t_rand2bd_xx_N14-negT}
shows~$\expec{\op\sigma_z}(t)$ for a ring-star geometry of the
environment as discussed in section~\ref{sec:spin-wheel-conf} and
demonstrated in Fig.~\ref{fig:sigma-z_t_rand2bd_xx}.  As expected,
the time averaged inversion is close now to~$0.6$ instead of~$-0.6$.
Thus for environments consisting of many spins, the regime~$T<0$ can
naturally be reached for sufficiently high energy in the environment.
In this way it is possible to study quantum thermodynamical effects at
negative absolute temperature and possibly even heat conduction
between embedded subsystems at temperatures of different sign.  This
``dual'' world, though artificial, may contribute to a better
understanding of thermal physics based on quantum mechanics.

\section{Conclusion}
\label{sec:conclusion}

For the example of a simple subsystem (spin~$1/2$) weakly coupled to a spin
environment we have shown that the spectral structure of the
environment has a strong influence on the decoherence of the central
system.  We have compared the decoherence of the central spin due to
random and structured coupling.

A noninteracting spin environment does not induce relaxation into the
local thermal state expected from the band structure, even though
energy is the only conserved quantity (the spectrum of the total
system is non degenerate).  In this case a majority of energy
eigenstates do not constitute linear combinations of all states
permitted by energy conservation and therefore the accessible part of
Hilbert space is considerably reduced for certain non equilibrium
initial states like the pure state~\eqref{eq:6}.

By introducing additional mutual coupling within the environment, thus
changing its spectrum, some of the properties of the completely random
perturbation with respect to relaxation of the central system are
restored, provided the coupling strength is adjusted appropriately.
By dynamically changing this coupling strength the thermalizing effect
of the environment on the central system should be adjustable.

We have concluded our investigation with some explorations on thermal
duality: Negative temperatures of the embedded system appear here as a
natural consequence of the spectral properties of the environment.

We thank the Landesstiftung Baden-W\"urttemberg for financial support.

\bibstyle{apsrev}
\bibliography{literatur}

\end{document}